%% file: paper.tex
\documentclass[sigconf]{acmart}
\renewcommand\footnotetextcopyrightpermission[1]{} 
\acmConference[ASE 2024]{IEEE/ACM Automated Software Engineering}{27 October - 1 November, 2024}{Sacramento, California, United States}

\usepackage{tcolorbox}
\usepackage{adjustbox}
\usepackage{enumitem}
\usepackage{tabularx}
\usepackage{array}
\usepackage{multirow}
\usepackage{soul}
\usepackage{breakurl}
\usepackage[htt]{hyphenat}

\usepackage{colortbl}%
\newcommand{\grayrow}{\rowcolor[gray]{0.925}}

\newcommand{\compactline}{\looseness=-1}

\usepackage{enumitem}
\setlist{itemsep=1pt,topsep=1pt,parsep=0pt,partopsep=0pt}

\newcommand{\tab}{\hspace*{1em}}

\AtBeginDocument{%
  \providecommand\BibTeX{{%
    \normalfont B\kern-0.5em{\scshape i\kern-0.25em b}\kern-0.8em\TeX}}}

\setcopyright{acmlicensed}
\copyrightyear{2018}
\acmYear{2018}
\acmDOI{XXXXXXX.XXXXXXX}

\begin{document}

\title{Semantic-Enhanced Indirect Call Analysis with Large Language Models}


\author{Baijun Cheng}
\affiliation{%
  \institution{Peking University}
  \city{Beijing}
  \country{China}}
\email{prophecheng@stu.pku.edu.cn}

\author{Cen Zhang}
\authornote{Corresponding authors}
\affiliation{%
  \institution{Nanyang Technological University}
  \city{Singapore}
  \country{Singapore}
}
\email{cen001@e.ntu.edu.sg}

\author{Kailong Wang}
\authornotemark[1]
\affiliation{%
 \institution{Huazhong University of Science and Technology}
 \city{Wuhan}
 \country{China}}
\email{wangkl@hust.edu.cn}

\author{Ling Shi}
\affiliation{%
  \institution{Nanyang Technological University}
  \city{Singapore}
  \country{Singapore}}
\email{ling.shi@ntu.edu.sg}

\author{Yang Liu}
\affiliation{%
  \institution{Nanyang Technological University}
  \city{Singapore}
  \country{Singapore}}
\email{yangliu@ntu.edu.sg}

\author{Haoyu Wang}
\affiliation{%
 \institution{Huazhong University of Science and Technology}
 \city{Wuhan}
 \country{China}}
\email{haoyuwang@hust.edu.cn}

\author{Yao Guo}
\authornotemark[1]
\affiliation{%
  \institution{Peking University}
  \city{Beijing}
  \country{China}}
\email{yaoguo@pku.edu.cn}

\author{Ding Li}
\affiliation{%
  \institution{Peking University}
  \city{Beijing}
  \country{China}}
\email{ding_li@pku.edu.cn}

\author{Xiangqun Chen}
\affiliation{%
  \institution{Peking University}
  \city{Beijing}
  \country{China}}
\email{cherry@sei.pku.edu.cn}

\renewcommand{\shortauthors}{Trovato and Tobin, et al.}

\begin{abstract}


In contemporary software development, the widespread use of indirect calls to achieve dynamic features poses challenges in constructing precise control flow graphs~(CFGs),
which further impacts the performance of downstream static analysis tasks.
To tackle this issue, various types of indirect call analyzers have been proposed.
However, they do not fully leverage the semantic information of the program, limiting their effectiveness in real-world scenarios.

To address these issues, this paper proposes Semantic-Enhanced Analysis~(SEA), a new approach to enhance the effectiveness of indirect call analysis.
Our fundamental insight is that for common programming practices, indirect calls often exhibit semantic similarity with their invoked targets. This semantic alignment serves as a supportive mechanism for static analysis techniques in filtering out false targets. 
Notably, contemporary large language models~(LLMs) are trained on extensive code corpora, encompassing tasks such as code summarization, making them well-suited for semantic analysis.
Specifically, SEA leverages LLMs to generate natural language summaries of both indirect calls and target functions from multiple perspectives.
Through further analysis of these summaries, SEA can determine their suitability as caller-callee pairs. 
Experimental results demonstrate that SEA can significantly enhance existing static analysis methods by producing more precise target sets for indirect calls.
\compactline

\end{abstract}

\maketitle

\input{sections/1.introduction-1}
\input{sections/2.motivation}
\input{sections/3.approach}
\input{sections/4.evaluation}

\input{sections/5.discussion}

\input{sections/6.related_and_conclusion}




\bibliographystyle{unsrt}
\bibliography{sections/references}

\end{document}

%% file: sections/1.introduction-1.tex
\section{Introduction}

Indirect calls~(\textit{\textbf{icalls}} hereafter) have been widely used in system and application software to enable dynamic behavior, simplify code structure, and improve code maintainability. They allow the target of a function call to be determined at runtime, providing better flexibility and modularity. As a result, they have become ubiquitous in large-scale software systems. For example, the Linux kernel 5.1.0 contains roughly 58K icalls, while Firefox has 37K icalls~\cite{MLTA}.

The use of icalls, with their runtime-dynamic nature, creates significant challenges for precise code analysis tools, such as those requiring an in-depth understanding of Interprocedural Control Flow Graph (ICFG) properties.
Typical examples of these scenarios are code navigation~\cite{SourceGraph} and code security analysis~\cite{SySeVR, VulExplainer, BeyondFidelity}.
The former, functionalities such as "Go to Definition" and "Find References", assist developers in locating identifier definitions and detecting callees.
However, icalls complicate these tasks, thereby reducing the effectiveness of existing tools.
The latter involves security analysis like taint analysis which is an ICFG-based method to identify whether a vulnerability is present or can be exploited, where an accurate icall resolution can provide significant benefits.
Therefore, having a precise and effective icall analysis is critical for comprehending program behavior and enabling enhanced code analysis capabilities.

Researchers have proposed to conduct icall analysis through pointer analysis~\cite{cpg, SVF, EmTaint} and type analysis~\cite{EFCFI, FineCFI, typro, TBCFI}. 
However, due to the inherent complexities of icalls (compared to regular function calls), it is hard to achieve high effectiveness using only interprocedural data flow or parameter types.
In contrast, we argue that \textit{high-level, e.g., natural language, textual semantics in code are also highly valuable for icall analysis}.
For example, in code repositories adhering to programming conventions, caller and callee functions (including icalls) often exhibit textual semantic similarities, which could be used to improve icall analysis results.

In recent years, the progress in AI~\cite{chang2024survey, shen2023boundary} and Natural Language Processing (NLP)~\cite{fanni2023natural, stahlberg2020neural, li2022end} has opened new opportunities to augment traditional icall analysis by extracting semantic data from the code.
Various software engineering applications are already making use of LLMs to better program analysis tasks~\cite{LLift,InferROI,Latte,gptscan,LLMDFA,KernelGPT,FuzzGPT,TitanFuzz}.
By incorporating LLMs, it is possible to establish natural language describable attributes, which are typically challenging to formalize.
This can potentially offer a superior understanding of code's semantics and context, which is comparable or even exceeding the grasp of human experts.

% ...

An intuitive perspective for utilizing LLMs for enhancing icall analysis is to boost code semantics interpretation, so we can better determine caller-callee relationships.
However, directly applying LLMs to this task presents two key challenges.
\textbf{Challenge \#1:}
The first challenge involves determining the key factors in or effective approaches to using LLMs to enhanced icall analysis.
It is necessary to refine the range of information extraction so it focuses on relevant functions.
Despite the various entities defined in the repositories, such as functions, structures, and unions, not all code pertains to icalls.
\textbf{Challenge \#2:}
The second challenge is how to effectively apply these approaches in complex, real-world projects.
The inherent intricacy of practical projects complicates the icall analysis.
Due to the largeness of code bases in practical projects, adaptive and meticulous methodologies are required to effectively gather the necessary information.
Further, as task complexity increases, relying solely on a single prompt might not be sufficient since the context surrounding caller-callee interactions may involve multiple entities.
Therefore, it is essential to develop strategies that can efficiently extract and use pertinent semantic information, mitigating the limitations of LLMs.
\compactline

\noindent\textbf{Our Work.} In this work, we present SEA (Semantic-Enhanced Analysis), a novel semantic-aware static analysis framework for icall analysis. Our approach aims to address the limitations of existing techniques by constructing semantic information within the context of callers and their target callees. 
With these semantic information, SEA can enable effective filtering of false targets, ultimately leading to more accurate and efficient icall analysis. 
For instance, a caller \texttt{module->create\_conf(cycle)} suggests that the corresponding callees are specifically designed to create configurations for particular modules. 
By leveraging this semantic similarity, SEA can effectively prune a significant portion of false targets.



To tackle \textbf{Challenge \#1}, considering the characteristics of the existing ICFG-based analysis and the textual nature of code semantics, we proposed a composite approach to utilize the code semantics for enhancement. Specifically, SEA first conducts traditional analyses~(e.g., type and pointer analyses), whose output is a set of candidate caller-callees pairs. It then apply code semantics analysis using LLM to further filter the candidates, modeling the problem as a binary classification task. 
This semantic-aware approach enables SEA to identify and filter out irrelevant functions, narrowing down the analysis scope to pertinent candidates.

To tackle \textbf{Challenge \#2}, we propose a two-step approach: semantic summary and caller-callee matching. First, we introduce a dedicated context database that extracts and collects necessary context information of callers and callees within the given project, inspired by observations of real-world programs and prior related work~\cite{autopruner, MLTA, IDECoder, cocomic} revealing semantic similarity between callers and callees. Our method employs an LLM-based extractor to comprehend the local and global contexts of the caller and callee, summarizing their task-specific characteristics and information, allowing SEA to gain a deeper understanding of their purpose and behavior within the broader context of the codebase.
Once the caller and callee summaries are generated, SEA utilizes LLMs to match them based on their semantic compatibility. Only when the caller's summary indicates that it can invoke the callee according to the extracted semantics, SEA will consider them a valid match. This semantic-driven matching process significantly reduces false positives and improves the accuracy of icall resolution.

We evaluate the effectiveness of SEA on 31 projects from oss-fuzz~\cite{ossfuzz}. 
The experimental results demonstrate that, in an environment without compilation options, 
SEA can improve the F1 score of the static analysis tool FLTA~\cite{FineCFI} by up to 24\% in the best-case scenario.
Additionally, in cases where only FLTA can perform the analysis, SEA improves its F1 score from 38\% to 67\%.
Compared to two more advanced tools MLTA~\cite{MLTA} and Kelp~\cite{Kelp}, SEA also demonstrate notable advantages in terms of flexibility and robustness. 
SEA is not restricted to analyzing callers that fit specific patterns, such as requiring the caller to be a simple function pointer or related to a non-escaped struct. 
The principles underlying SEA are applicable to a wide range of indirect calls. 

\noindent\textbf{Contributions.} Our contributions are summarized as follows:

\begin{itemize}[leftmargin=8pt]

\item We propose SEA, the first semantic-enhanced approach for indirect call analysis with the help of LLMs. 
Compare to conventional techniques~\cite{MLTA, Kelp, SVF},
our approach can fully leverage the textual semantics in code to better determine icall relationships. 

\item We implement SEA and conduct experiments by integrating SEA with existing conventional static analysis methods. Results demonstrate that SEA can accurately filter out invalid caller-callee pairs, and improve existing static analysis significantly.

\item We curate an extensive benchmark dataset sourced from real-world programs, leveraging fuzzing techniques~\cite{ossfuzz} in its construction. To facilitate the future research on this topic, we release our artifacts at~\cite{CodeAnalyzer}.

\end{itemize}

%% file: sections/2.motivation.tex
\vspace{-5pt}
\section{Preliminaries}

\subsection{Background}





\begin{figure}[t]
  \centering
  \includegraphics[width=0.2\textwidth]{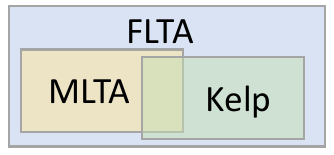}\vspace{-0.4cm}
  \caption{Relationship between traditional methods.}
  \vspace{-2mm}
\label{fig:traditional_method}\vspace{-0.3cm}
\end{figure}

\textbf{Existing Work on Indirect Call Analysis}
Currently, scalable static analysis methods include FLTA~\cite{FineCFI}, MLTA~\cite{MLTA}, and Kelp~\cite{Kelp}.
Their relationships are depicted in Figure~\ref{fig:traditional_method}, highlighting their different application scenarios. FLTA is based on simple function signature matching, making it suitable for analyzing all icalls. 
MLTA~\cite{MLTA} proposes a hierarchical analysis approach based on struct type levels to enhance the precision of FLTA, which is effective only when function pointers are encapsulated within structs and the corresponding structs do not exhibit escape behavior. 
Kelp~\cite{Kelp} introduces a regional pointer analysis technique tailored for scenarios characterized by simple data flow patterns.
When icalls do not satisfy the specified patterns, MLTA and Kelp would fall back to other analysis algorithms (MLTA falls back to FLTA, while Kelp falls back to either MLTA or FLTA).
An intuitive solution is to augment the analytical rules to encompass as many cases as possible. However, such an approach entails the precision-recall balance dilemma, wherein the endeavor to prune a greater number of false targets may inadvertently leads to the pruning of a portion of true targets.
During our preliminary study on the benchmarks, among all successfully parsed and analyzed icalls, we observed that Kelp figured out 10.4\% of cases, while MLTA covered 46.3\% of cases. FLTA handled the remaining 43.3\% of cases. 
This leaves significant room for further optimization.

\noindent\textbf{Underutilization of Code Semantics.} Existing static analysis methods lack comprehension of code semantics.
Typically, within code repositories adhering to programming standards, function pointers and target functions often exhibit semantic similarities superficially. 
Thanh \textit{et al.}~\cite{autopruner} proved that semantic features are helpful in call graph pruning.
However, existing static analysis methods  capture this aspect inadequately.
We observe that in many cases not covered by Kelp, there exists a strong semantic similarity between icalls and their invoked targets. 
Leveraging these features, we can further refine the target sets.

\begin{figure*}[t]
  \centering
  \includegraphics[width=0.95\textwidth]{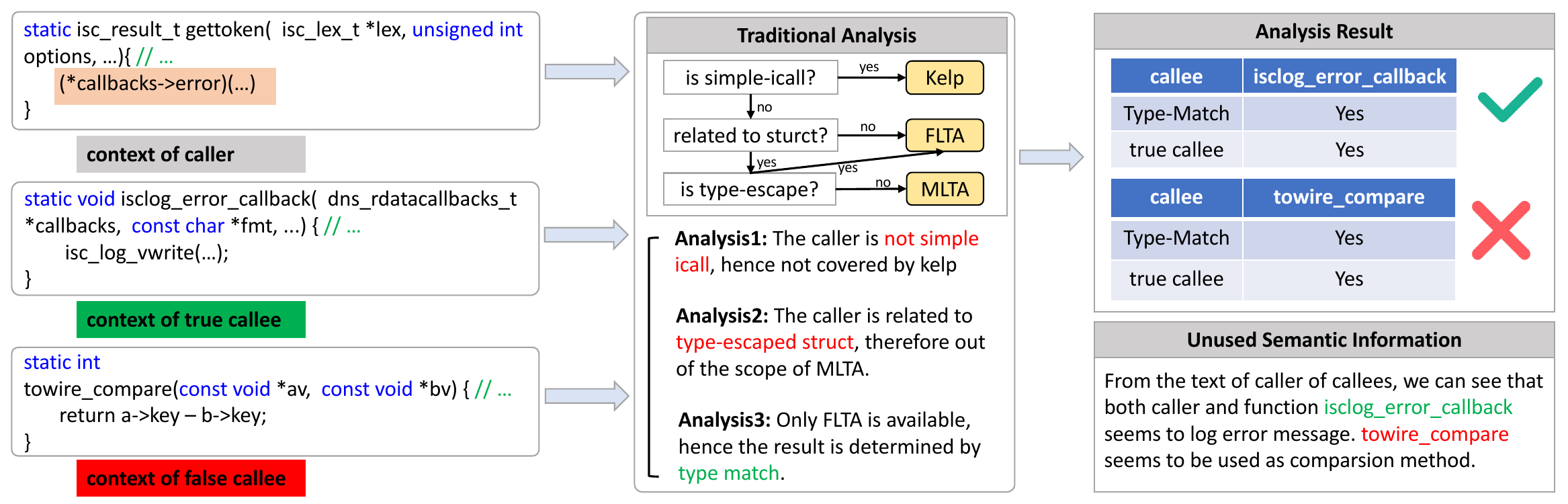}
  \vspace{-4mm}
  \caption{Example indirect-call and target functions.}
  \vspace{-2mm}
  \label{fig:icall_example}
\end{figure*}

\noindent\textbf{Capabilities of LLMs.}  Recently, significant advancements have been made in software analysis by applying LLMs. Previous studies~\cite{LLift, TitanFuzz, InferROI, pei2023can, KernelGPT} have demonstrated the superiority of LLMs over traditional program analysis methods in summarizing path conditions, seed variation, source-sink analysis, summarizing loop invariants, and extract critical attributes.
In our scenario, we find that LLMs could play a significant role in code semantic analysis. 
This is attributed to LLMs being trained on extensive code corpora, encompassing tasks such as code translation, code summarization, document generation, etc. 
Such training endows LLMs with strong capabilities in comprehending code semantics.

Despite offering new perspectives for software analysis, LLMs face several challenges~\cite{li2024digger,li2024lockpicking,deng2024pandora,wang2024metmap,li2024glitch,liu2023prompt,wang_ndss_2024}. Firstly, their limited context capacity means that we cannot input large chunks of code for analysis at once. 
Consequently, it may be impossible to provide all relevant context for both the caller and callee in a single input to LLMs. 
Secondly, LLMs are prone to generating incorrect responses, a phenomenon known as hallucination~\cite{wang2024oopsla, Hallucination, FightFire}.
Lastly, their performance deteriorates as task complexity increases~\cite{DecomposedPrompt}, and even the chain-of-thought strategy may become ineffective~\cite{CoT}.

\subsection{Motivating Example}
Figure~\ref{fig:icall_example} shows the general process of traditional static analysis methods on an icall example from the bind9~\cite{bind9} project. 
The caller has two potential callees: the true target \texttt{isc\_log\_error\_callback}, and the false target \texttt{towire\_compare}.
The analysis falls outside the scope of Kelp~\cite{Kelp} as it does not meet the definition of a simple icall according to the tool's criteria. 
Furthermore, despite being related to a struct, the challenge of type-escape~\cite{MLTA} renders it unanalyzable by another SOTA approach MLTA~\cite{MLTA}. 
Consequently, FLTA~\cite{FineCFI} stands as the sole capable analyzer in this scenario.
The analysis outcome of FLTA marks both callees as true targets, considering the casting relationship between void* and other pointer types.
\compactline

Reviewing the case discussed above, we observe that current static analysis methods overlook the intrinsic semantic information of the program. 
This semantic information is essential for enhancing the accuracy of the analysis results.
For instance, by comparing the context of the true callee and the caller, we can see that their functions are related to reporting errors, whereas the false callee, based on textual analysis, is a comparison function that is unlikely to be invoked by the caller.

%% file: sections/3.approach.tex
\section{Methodology}

\begin{figure*}[t]
  \centering
  \includegraphics[width=0.95\textwidth]{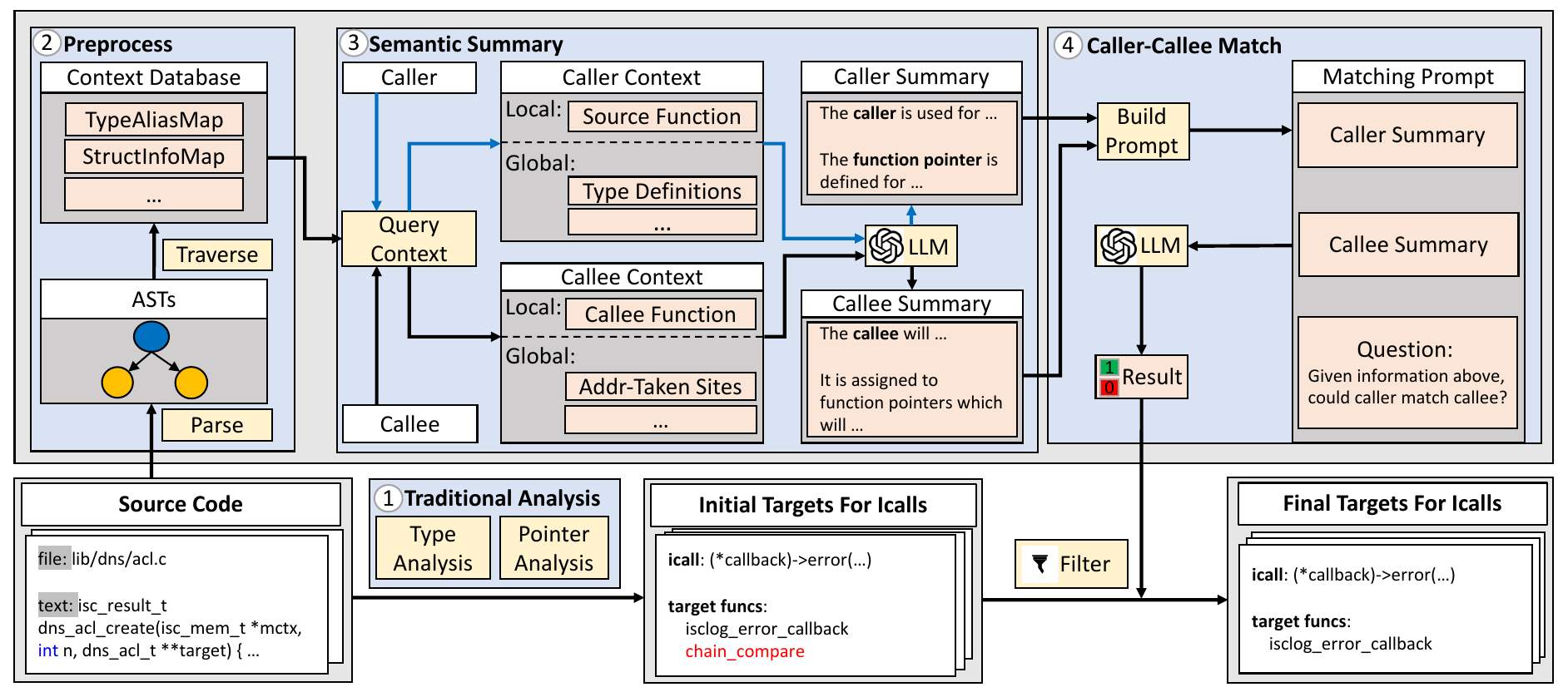}
  \vspace*{-2mm}
  \caption{Overview of SEA.}
  \label{fig:overview}
\end{figure*}

Figure~\ref{fig:overview} presents the overview of SEA, which is inspired by a call graph pruning tool Autopruner~\cite{autopruner}. 
It begins by utilizing traditional static analysis~(\S~\ref{sec:static_analysis}) to ascertain a superset of the target set for each icall. 
Subsequently, it harnesses the semantic comprehension capabilities of a language model to refine this superset through the binary classification of caller-callee pairs.

During semantic analysis, SEA initially preprocesses the source code of a project, constructing a context database that includes crucial context information such as \texttt{TypeAliasMap}~(stage 2, \S~\ref{sec:preprocess}). 
Subsequently, SEA queries corresponding context from the context database based on both caller and callee, and inputs it to LLM, prompting it to analyze the purposes of the caller and callee and generate natural language summaries~(stage 3, \S~\ref{sec:semantic_summary}). 
Finally, SEA constructs the ultimate matching prompt based on the summaries of the caller and callee, which is then analyzed by LLM to determine if they form a caller-callee pair. 
Based on the analysis by LLM, SEA refines the initial target set into the final target set~(stage 4, \S~\ref{sec:caller_callee_match}).

\subsection{Traditional Analysis}~\label{sec:static_analysis}


The first step of SEA involves identifying an initial target function set for each icall from all functions of a project.
This target set theoretically encompasses numerous false positives and necessitates further refinement. 
Currently, FLTA~\cite{FineCFI}, MLTA~\cite{MLTA}, and kelp~\cite{Kelp} can achieve this objective.
However, the evaluation of these three static analysis methods is predominantly conducted within the LLVM IR context, with no current implementation at the source code level. 
At the source code level, our implementations largely follow the analysis algorithms used at the IR level. The additional components that need to be implemented include (1) the identification of address-taken functions and (2) the strategy of type matching.
\compactline

\noindent\textbf{Identification of Address-Taken Functions.} A presumption in icall analysis is that the target of an icall must be an address-taken function, meaning a function whose address is assigned to a pointer variable. 
In LLVM IR, identifying address-taken functions is straightforward as users can easily determine if a function's address is taken by utilizing LLVM's API~\cite{LLVM}.
However, this process is not as straightforward at the source code level, especially in the absence of compilation options. 
We can solely rely on analyzing the abstract syntax trees~(ASTs) to identify the set of address-taken functions within a project.
The traversal rule is defined as follows:

\begin{itemize}[leftmargin=8pt]
\item For each identifier node \texttt{id}, if \texttt{id} is not the first child of a call expression, also \texttt{id} is not a global or local variable name, we add \texttt{id} to address-taken function name set.  
\end{itemize}


\noindent\textbf{Strategy of Type Matching.} Before type matching, we must identify all types of call arguments and function parameters.
This step could be done by traversing ASTs.
However, during this process, we may encounter parsing errors. 
For instance, the presence of macros such as \texttt{DNS\_\_DB\_FLARG} in variable definition statements like \texttt{dns\_rdataset\_t *rdataset DNS\_\_FLARG} can cause parsing errors, preventing us from identifying the type of the variable \texttt{rdataset}. 
In such cases, we mark these variables with unresolved parsing errors as unknown types.
Consequently, our type-matching algorithm differs from that of K. Lu et al.~\cite{MLTA}. 
Besides, our approach also considers factors such as variable parameters, leading us to employ a more conservative matching strategy. 
Specifically, our type-matching approach differs from MLTA in the following ways:

\begin{itemize}[leftmargin=8pt,  parsep=1pt]

\item Our approach does not rely on matching function signatures. Instead, it compares the types of each parameter individually. This comparison is conducted under the condition that the number of parameters aligns, accounting for variadic parameters when necessary.

\item Parameters marked as unknown types are treated conservatively, being assumed to match with all types.

\item We establish that \texttt{void*} and \texttt{char*} can be matched with any pointer type, as observed in benchmarks where similar type conversions exist.

\end{itemize}

This conservative strategy may introduce more false targets but also preserves a greater number of true targets.

\begin{figure}[t]
  \centering
  \includegraphics[width=0.47\textwidth]{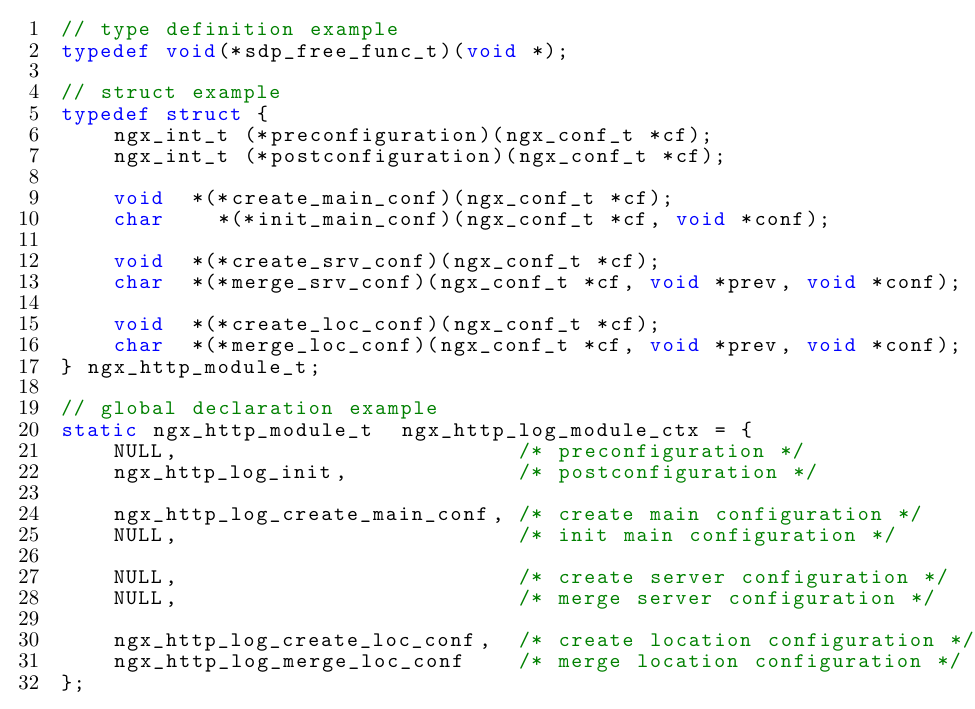}
  \vspace{-4mm}
  \caption{Example of global context.}
  \vspace{-4mm}
  \label{fig:struct_example}
\end{figure}


\subsection{Preprocess}~\label{sec:preprocess}


The primary objective of this step is to construct a context database for subsequent analysis, akin to the context retrieval methods employed in IDECoder~\cite{IDECoder} and Cocomic~\cite{cocomic}. 
However, our work distinguishes itself by concentrating on obtaining context related to the caller and callee for caller-callee pair matching. 
In contrast, IDECoder and Cocomic focus on extracting other textual information that the current code depends on to facilitate code completion.

To identify the optimal context for input into an LLM for analysis, we reference other works. 
Autopruner~\cite{autopruner}, in particular, demonstrated the critical role of both caller and callee functions in the semantic analysis of call graphs.
The text of these functions is typically located in the local code segments of the caller and callee, which we refer to as the local context.
Additionally, beyond the local context, there are code segments semantically related to the caller and callee, which we refer to as the global context.
Compared to the local context, we observe that the global context provides a more high-level description of the caller and other functions of the same type.
Through manual analysis of several projects, we identify three types of global context that are beneficial for understanding the semantics of caller and callee functions: type definitions, global variable declarations, and struct definitions.
We illustrate this with Figure~\ref{fig:struct_example}, in which the type definition statement defines the \texttt{sdp\_free\_func\_t} function type, indicating that this class of function pointers is related to freeing memory.
Where in struct \texttt{ngx\_http\_module\_t}, eight different function pointers are defined, each corresponding to a handler in the nginx http module, which needs to perform different functions. 
In the subsequent global variable declaration, an nginx module \texttt{ngx\_http\_log\_module\_ctx} is defined, where functions such as \texttt{ngx\_http\_log\_create\_main\_conf} and \texttt{ngx\_http\_log\_init} are assigned to corresponding fields. 
When the caller's function pointer or the callee's address-taken site involves the above type, struct, or global variable, incorporating their text into the prompt can better facilitate the LLM's understanding of its semantics.

\begin{figure}[t]
  \centering
\includegraphics[width=0.47\textwidth]{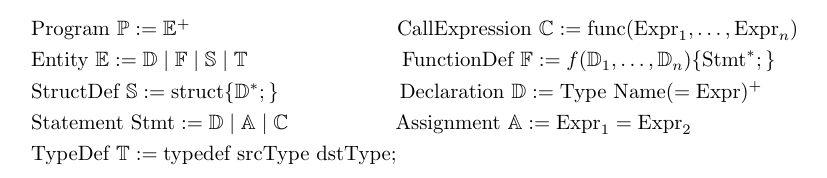}
  \vspace{-5mm}
  \caption{Abstraction of language.
  }
  \vspace{-5mm}
  \label{fig:language}
\end{figure}

In addition, the callee's address-taken site primarily originates from three sources: assignment expressions, initializers of declarations, and call arguments.
Therefore, we abstract the language we process into the form illustrated in Figure~\ref{fig:language}, where a program is viewed as a collection of entities. 
Each entity can be a global variable declaration, function, type, or struct definition. 
Declarations can specifically refer to global variables, local variables, fields, or parameter declarations. Regarding statements, we focus on call, assignment, and declaration statements, as these may indicate address-taken sites of a function.
When traversing the AST, we use seven maps to store relevant information: \texttt{TypeAliasMap}, \texttt{StructInfoMap},
\texttt{GlobalVarMap},
\texttt{FunctionMap}, \sloppy\texttt{FuncNameToCallExprsMap}, \texttt{FuncNameToDeclarationsMap}, and \texttt{FuncNameToAssignmentsMap}. 
The \texttt{FunctionMap} stores basic information about a function, including details about its parameters and local variables. 
The \texttt{TypeAliasMap}, \texttt{StructInfoMap}, and \texttt{GlobalVariableMap} store information related to type aliases, defined structures, and declared global variables, respectively. 
Furthermore, the \texttt{FuncNameToCallExprsMap}, \texttt{FuncNameToDeclarationsMap}, and \texttt{FuncNameToAssignmentsMap} store basic information about each function's call expression, variable declaration, and assignment statement address-taken sites, respectively.
The corresponding rules can be described as follows:

\begin{itemize}[leftmargin=8pt, parsep=1pt]

\item For each \texttt{TypeDef} statement, \texttt{srcType} represents the existing type name, which may be a function type, and \texttt{dstType} represents the new type name. 
We store the information of \texttt{srcType} and \texttt{dstType} to the \texttt{TypeAliasMap}.

\item For each \texttt{StructDef}, which contains a series of field declarations, some of these fields may be function pointers. 
We store the struct name along with the corresponding field information in the  \texttt{StructInfoMap}.

\item For each \texttt{Declaration D}, if it declares a global variable, we add variable name and \texttt{D} to \texttt{GlobalVarMap}.

\item For each \texttt{FunctionDef} statement, we parse its declarator to obtain the type and name of each parameter and analyze the function body to retrieve information about local variables. 
We then add this information to the \texttt{FunctionMap}.

\item For each \texttt{CallExpression C} , if its $i$-th argument expression $Expr_i$ references an address-taken function \texttt{func}, we add \texttt{C} and the argument index $i$ to \texttt{FuncNameToCallExprsMap[func]}.

\item For each \texttt{Assignment A}, if the right-hand side expression $Expr_2$ references an address-taken function \texttt{func}, we add \texttt{A} to \texttt{FuncNameToAssignmentsMap[func]}.

\item For each global or local variable \texttt{Declaration D}, if it includes an initializer that references an address-taken function \texttt{func}, we add \texttt{D} to \texttt{FuncNameToDeclarationsMap[func]}.

\end{itemize}

\subsection{Semantic Summary}~\label{sec:semantic_summary}


\begin{figure}[t]
  \centering
\includegraphics[width=0.5\textwidth]{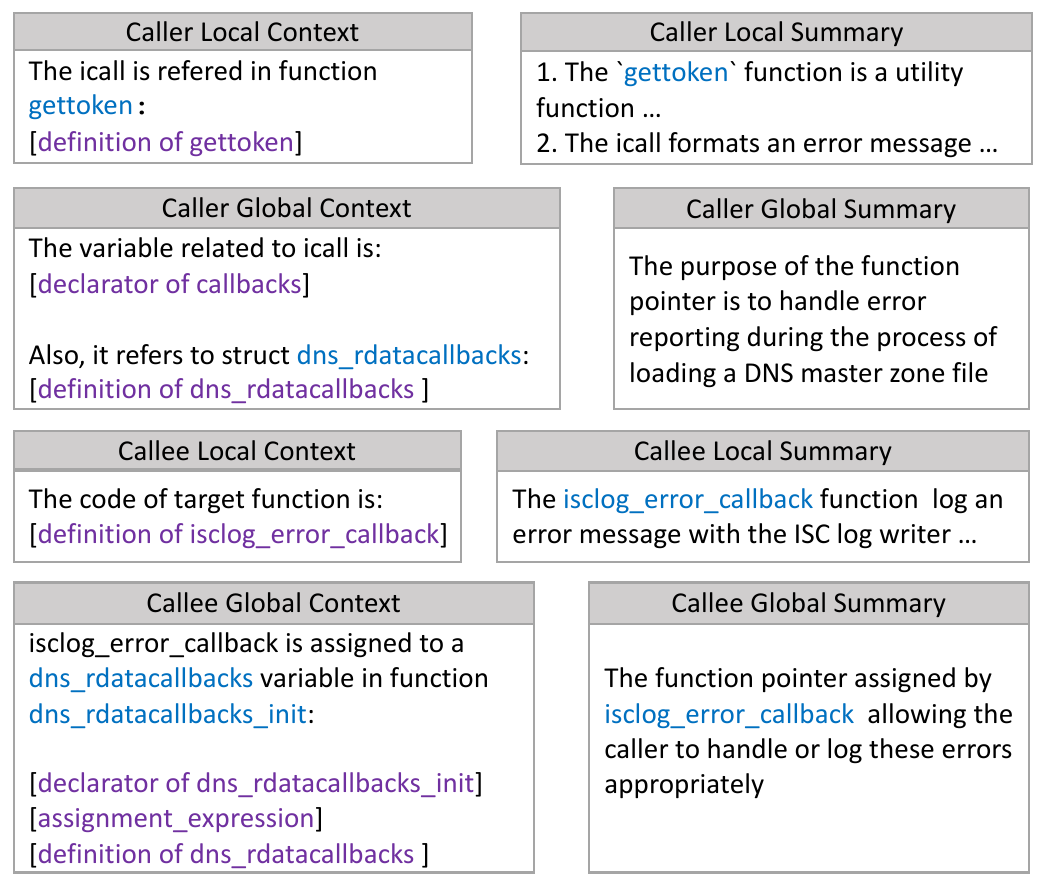}
  \vspace{-6mm}
  \caption{Example of context for caller and callee.
  }
  \vspace{-4mm}
  \label{fig:context_example}
\end{figure}

\vspace{-2mm}

In the previous section~(\S~\ref{sec:preprocess}), we discuss the global and local contexts that are useful for caller-callee matching and the context database used to store them. 
Subsequently, the primary objective of this step is to extract the corresponding local and global contexts for a given caller and callee from the context database and input them into the LLM for analysis.
We follow a four-step process to obtain the caller's and callee's local and global contexts. 
The LLM then summarizes these contexts, generating natural language descriptions of their purposes, which we refer to as the caller local summary, caller global summary, callee local summary, and callee global summary, respectively. 
It is important to note that if no global context is collected, we skip the corresponding step. 
Finally, we input all the summaries generated by the LLM into the LLM again to perform the caller-callee matching.
As the example shown in Figure~\ref{fig:context_example}, SEA successfully extracts the contexts of the caller and callee, generates the corresponding natural language summaries, and further combines these four summaries through the LLM to successfully identify the caller-callee pair.

\noindent\textbf{Query Context.} The process of extracting the local context for the caller and callee is straightforward: we simply retrieve the corresponding function text from the \texttt{FunctionMap}. 
Extracting the caller's global context is slightly more complex. 
We need to parse the caller's corresponding AST node and analyze its function pointers. 
The rules for this analysis can be broadly described as follows:

\begin{itemize}[leftmargin=8pt, parsep=1pt]

\item We extract the corresponding variable declaration from the \texttt{GlobalVarMap} if the function pointer is a global variable, or from \texttt{FunctionMap} if it is a local variable, and add it to the global context.

\item If the corresponding function pointer is a struct field, we extract the corresponding struct text from the \texttt{StructInfoMap} and add it to the global context.

\item If the function pointer variable's declaration involves a type alias, we extract the corresponding type declaration text from the \texttt{TypeAliasMap} and add it to the global context.

\end{itemize}

In comparison to extracting the caller's global context, extracting the callee's global context is even more complex. 
While extracting the caller's global context only requires analyzing the caller's function pointers, extracting the callee's global context involves analyzing its corresponding address-taken sites, which may be initializers, assignments, or call arguments. 
Additionally, a callee may contain multiple address-taken sites.
For the callee function \texttt{func}, we first extract the corresponding global context for each address-taken site according to the following rules:

\begin{itemize}[leftmargin=8pt, parsep=1pt]\label{itemize:caller_global}

\item For assignment site in  \texttt{FuncNameToAssignmentsMap[func]} and \texttt{FuncNameToDeclarationsMap[func]}~(initializer and assignment share similar process), we parse the left expression of the assignment statement, which is the function pointer being assigned, according to the method described in the previous step of extracting caller global context. 
We then combine the corresponding global context with the assignment site text to form the global context for the current assignment site.

\item For each call expression in \texttt{FuncNameToCallExpr[func]}, we obtain its call-chain with respect to \texttt{func}, represented as \texttt{[($call_1$, $target_1$), ..., ($call_n$, $target_n$)]}. 
Here, $target_i$ is the target of $call_i$, and \texttt{func} is used as an argument in one of the calls. 
The call chain's endpoint is either another icall or a point where \texttt{func} is assigned to a variable or directly invoked in $target_n$. 
If it is ultimately assigned to a variable, we analyze the corresponding assignment or initializer following the previous steps to obtain the global context. 
We then combine this with the call-chain text (where each ($call_i$, $target_i$) consists of the text of the corresponding call statement and the function declarator text) to form the global context for the address-taken site.
If it is directly invoked, we combine the call statement with the call-chain text to form the final global context.

\end{itemize}

\noindent\textbf{Context Summary.} A critical aspect of this process is prompt design. 
To date, various prompt strategies~\cite{PromptEngineerSurvey} have been proposed to enhance the performance of LLMs. 
Among these, task decomposition~\cite{DecomposedPrompt, LLift} and chain-of-thought~\cite{CoT} have proven useful in handling complex tasks or when dealing with lengthy input sequences. 
Considering that combining the local and global contexts of the caller and callee may exceed the context length, and given that these contexts describe their specific functionalities and the roles these function pointers are meant to fulfill, with semantic differences between them.
We adopt the task decomposition strategy. 
Specifically, we decompose the caller-callee matching task into two steps: semantic summarization and matching.
Our intuition is that semantic summarization can distill the critical information from the caller's and callee's contexts into a concise natural language summary. 
This approach not only reduces the likelihood of excessively long inputs but also minimizes the inclusion of irrelevant information during the matching process.
Semantic summarization consists of four steps, and each step uses a prompt to process the caller's or callee's global or local context and generate the corresponding natural language summary.
Each prompt consists of two parts: \texttt{[context]} and \texttt{[instruction]}. 
The \texttt{[context]} is the relevant code snippet retrieved from the context database, while the \texttt{[instruction]} is a sentence that guides the LLM in analyzing the context. 
\compactline

\vspace{-3mm}
\begin{center}
\fcolorbox{black}{gray!10}{\parbox{\linewidth}{The local context for the indirect-call is listed as follows:

[local context]

Analyze the functionality of the indirect call and response with a concise summary.}}
\end{center}

Note that a callee function may have multiple address-taken sites, and each site may be assigned to variables with different semantics, resulting in different summaries. Therefore, when multiple address-taken sites are present, we use a single prompt to combine the summaries of all address-taken sites, as follows:
\compactline

\vspace{-3mm}

\begin{center}
\fcolorbox{black}{gray!10}{\parbox{\linewidth}{The summaries of each address-taken site for the target function are:

[summary1],

...

[summaryn]

Please consolidate those summaries.}}
\end{center}

\subsection{Caller-Callee Match}~\label{sec:caller_callee_match}

After obtaining the summaries for the caller and callee, we instruct the LLM to determine if there is a calling relationship between them using the following prompt:

\vspace{-3mm}

\begin{center}
\fcolorbox{black}{gray!10}{\parbox{\linewidth}{
The subsequent text provides the summary of the caller and callee:

\# 1.summary of caller

\#\# 1.1.[CallerLocalSummary]

\#\# 1.2.[CallerGlobalSummary]

\# 2.summary of callee

\#\# 2.1.[CalleeLocalSummary]

\#\# 2.2.[CalleeGlobalSummary]

Assess if the caller could invoke callee based on the semantic information given above. Answer with `yes' or `no'.
}}
\end{center}

In this step, the LLM determines whether a real calling relationship exists between the caller and each potential callee. 
After obtaining the LLM's analysis results, we use them to refine the outcomes of traditional static analysis methods.

%% file: sections/4.evaluation.tex
\section{Evaluation}

\subsection{Experimental Settings}

\begin{table}[t]
\centering
\caption{Basic information of benchmarks. IN denotes the number of icalls, and VN denotes the number of valid icalls. SLoc denotes the number of lines of code}\vspace{-0.2cm}
\begin{adjustbox}{width=0.47\textwidth}
\resizebox{0.8\linewidth}{!}{
\begin{tabular}{l c c c || l c c c}
\hline
\textbf{Project} & \textbf{IN} & \textbf{VN} & \textbf{SLoc} & \textbf{Project} & \textbf{IN} & \textbf{VN} & \textbf{SLoc} \\
\hline
\grayrow
bind9 & 46 & 45 & 388k & bluez & 4 & 4 & 450k \\
cairo & 41 & 41 & 242k & cyclonedds & 47 & 47 & 286k \\
\grayrow
dovecot & 60 & 39 & 486k & fwupd & 31 & 28 & 299k \\
gdbm & 18 & 18 & 18k & gdk-pixbuf & 3 & 3 & 88k \\
\grayrow
hdf5 & 94 & 89 & 1365k & igraph & 19 & 6 & 275k \\
krb5 & 18 & 7 & 409k & libdwarf & 202 & 17 & 154k \\
\grayrow
libjpeg-turbo & 462 & 209 & 142k & libpg\_query & 11 & 11 & 520k \\
libsndfile & 8 & 8 & 67k & libssh & 15 & 15 & 96k \\
\grayrow
librabbitmq & 6 & 6 & 13k & lua & 7 & 2 & 33k \\
lxc & 7 & 7 & 74k & md4c & 39 & 6 & 23k \\
\grayrow
mdbtools & 2 & 2 & 18k & nginx & 23 & 23 & 629k \\
opensips & 8 & 8 & 603k & oniguruma & 268 & 18 & 95k \\
\grayrow
pjsip & 43 & 22 & 809k & postfix & 4 & 3 & 223k \\
rtpproxy & 161 & 8 & 136k & selinux & 21 & 21 & 518k \\
\grayrow
sudo & 96 & 95 & 282k & tmux & 11 & 11 & 73k \\
vlc & 1065 & 31 & 2560k & & & & \\
\hline
\end{tabular}
}
\end{adjustbox}\vspace{-0.4cm}
\label{tab:benchmark}
\end{table}

\noindent
\textbf{Benchmark Selection} \tab
We select test programs from the oss-fuzz projects~\cite{ossfuzz} and follow their build scripts to download, compile, and dynamically run them with a fuzz testing tool to generate ground-truth data. 
During compilation, we implement a custom LLVM pass to instrument the programs, enabling the dumping of information related to the ground truth when indirect calls are executed. 
We used AFL++~\cite{AFLplusplus} as the fuzz testing tool. 
Due to poor version management of many projects under oss-fuzz, some projects failed to compile successfully and are excluded from our benchmark.
Additionally, we find that some programs either produced invalid results due to parsing errors or achieved 100\% F1 scores using only FLTA, leading us to exclude these projects as well. 
Ultimately, we collected 31 projects for our benchmark. Detailed information is listed in Table~\ref{tab:benchmark}, which includes the number of icalls presented in the ground-truth dataset.
However, many icalls are either wrapped in macros or sufferred from other parsing errors, resulting in the absence of valid analysis results. 
The valid icall number~(VN) indicates the icalls that can be correctly analyzed. 
The fundamental solution to these issues is to preprocess the project with a compiler before analyzing the preprocessed code. 
However, this conflicts with the preset condition of not using compilation options.
\compactline

\noindent
\textbf{LLM selection} \tab
Our experiment requires a substantial token expenditure, rendering the use of SOTA close-source models~\cite{ChatGPT, Claude, gemini} costly. 
Consequently, we primarily focused on leveraging open-source LLMs.
Regarding model selection, we focused on instruction-tuned~\cite{InstructTune} models. 
Based on the intuition that larger models tend to perform better, we chose Qwen1.5-72B-Chat~\cite{qwen}\footnote{At the time of our experiment, Qwen1.5-110B had not yet been released. Additionally, the deployment of the 110B large model was beyond our capabilities.} and LLaMA3-70B-Instruct~\cite{llama3}~(\textbf{\textit{qwen}} and \textbf{\textit{llama3}} hereafter) for experiments.
\compactline

\begin{figure}[t]
  \centering
  \includegraphics[width=0.5\textwidth]{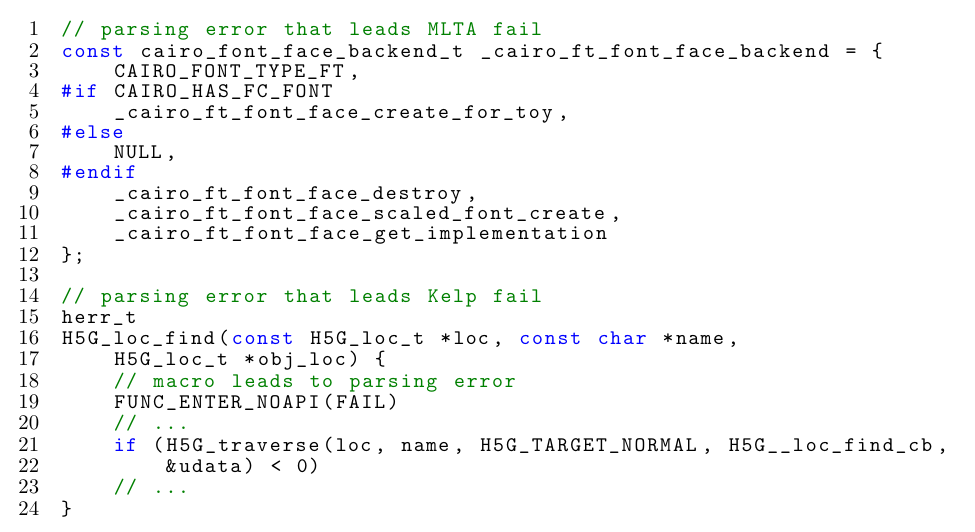}
  \vspace{-6mm}
  \caption{Parsing errors introduced by conditional compilation and macro.}
  \vspace{-4mm}
  \label{fig:error_example}
\end{figure}

\noindent
\textbf{Baseline Approach} \tab
We consider three static analysis methods—FLTA~\cite{FineCFI}, MLTA~\cite{MLTA}, and Kelp~\cite{Kelp}—as baselines. 
At the LLVM IR level, MLTA optimizes FLTA's results in some cases, and Kelp further optimizes MLTA's results. 
From FLTA to Kelp, precision improves without introducing false negatives. Therefore, we only need to compare with Kelp. 
However, at the source code level, the situation is more complex. 
Language features such as conditional compilation and macros can cause parsing errors in the code before preprocessing, leading to incorrect fact propagation and resulting in false negatives.
Figure~\ref{fig:error_example} illustrates one example, where the presence of the conditional macro \texttt{CAIRO\_HAS\_FC\_FONT} in the initializer of the \texttt{\_cairo\_ft\_font\_face\_backend} variable causes errors in AST analysis. 
This results in the function addresses like \texttt{\_cairo\_ft\_font\_face\_create\_for\_toy} being incorrectly associated with other fields in the \texttt{cairo\_font\_face\_backend\_t} structure, leading to false negatives in MLTA. 
Similarly, in the \texttt{H5G\_loc\_find function}, the macro \texttt{FUNC\_NOAPI(FAIL)} causes parsing errors in the AST beyond the function body, preventing the correct analysis of the \texttt{H5G\_traverse} call. 
Consequently, \texttt{H5G\_\_loc\_find\_cb} is not propagated to its corresponding simple function pointer, resulting in false negatives in Kelp.
The fundamental solution to these issues remains to preprocess the project with a compiler before analyzing the preprocessed code.

\begin{table}[t]
    \centering
     \caption{Binary classification result for SEA with qwen at temperature 0.5.  \textbf{baseline} refers to treating all call edges analyzed by the baseline approach as true targets.}
     \vspace{-0.2cm}
\resizebox{1.0\linewidth}{!}{
    \begin{tabular}{c|c|c|c|c|c|c|c|c|c|c|c|c}
     \hline
     \multirow{2}{*}{\textbf{Approach}} & \multicolumn{4}{c}{FLTA} & \multicolumn{4}{c}{MLTA} & \multicolumn{4}{c}{Kelp} \\
     \cline{2-13}
     & A & P & R & F & A & P & R & F & A & P & R & F \\
     \hline
     \textbf{baseline} & 17.3 & 17.3 & 100.0 & 27.3 & 26.4 & 26.4 & 100.0 & 38.2 & 31.4 & 31.4 & 100.0 & 43.4\\
     \hline
     \grayrow
     \textbf{SEA} & 76.0 & 41.2 & 98.9 & \underline{\textbf{55.3}}
     & 66.5 & 41.1 & 98.9 & \underline{\textbf{55.3}}
     & 64.8 & 44.2 & 99.4 & \underline{\textbf{57.9}} \\
     \hline
    \end{tabular}
}
    \label{tab:binary_result}
    \vspace{-1mm}
\end{table}

\noindent
\textbf{SEA Setup} \tab We assumed that the caller and all its callees reside within the same scope. We define this scope as being in the same directory and its subdirectories as the caller's source file.

\noindent
\textbf{Evaluation Metrics} \tab
Three evaluated metrics, \textit{i.e.}, precision~(P), recall~(R), and F1~(F) are defined as following.
For each icall, $C^{pr}$ is the callee set predicted by SEA for a given icall, $C^{gt}$ represents the corresponding callee set in the ground truth.

\begin{equation}
    P = \frac{|C^{pr} \cap C^{gt}|}{|C^{pr}|} \quad
    R = \frac{|C^{pr} \cap C^{gt}|}{|C^{gt}|} \quad
    F = \frac{2 \cdot P \cdot R}{P + R}
\end{equation}

\noindent
\textbf{Research Questions~(RQs)} \tab
Our key RQs are as following:

\begin{itemize}[leftmargin=8pt, parsep=1pt]
\item \textbf{RQ1:} Is SEA effective in enhancing the performance of current static analysis?

\item \textbf{RQ2:} To what extent do the global and local context of caller and callee contribute to SEA?
\end{itemize}

\begin{table*}[t]
    \centering
     \caption{icall analysis results for different setups: FLTA, MLTA, and Kelp denote the full evaluation, while FLTA-exclusive and MLTA-exclusive denote exclusive evaluations.}\vspace{-0.2cm}
\resizebox{0.8\linewidth}{!}{
    \begin{tabular}{c|c|c|c|c|c|c|c|c|c|c|c|c|c|c|c|c}
        \hline
        \multirow{2}{*}{\textbf{Model}} & \multirow{2}{*}{\textbf{Temperature}} &
         \multicolumn{3}{c}{\textbf{FLTA}} & \multicolumn{3}{c}{\textbf{MLTA}} & \multicolumn{3}{c}{\textbf{Kelp}} &
         \multicolumn{3}{c}{\textbf{FLTA-exclusive}} &
         \multicolumn{3}{c}{\textbf{MLTA-exclusive}} \\ \cline{3-17}
         & & P & R & F & P & R & F & P & R & F & P & R & F & P & R & F\\
         \hline
        \hline
        \multirow{3}{*}{\textbf{qwen}} & 0 & 48.7 & 96.9 & 59.0 & 52.0 & 96.5 & 61.3 & 54.9 & 96.6 & 63.7 & 59.3 & 97.5 & 66.4 & 52.9 & 98.0 & 61.6 \\ \cline{2-17}
        
        & 0.25 & 48.8 & 97.3 & 59.1 & 51.9 & 96.8 & 61.4 & 54.8 & 96.9 & 63.7 & 59.4 & 97.7 & 66.4 & 52.9 & 98.0 & 61.6  \\ \cline{2-17}

        \grayrow \cellcolor{white}
        & 0.5 & 49.1 & 97.3 & \underline{\textbf{59.4}} & 51.9 & 96.9 & 61.4 & 55.0 & 97.0 & \underline{\textbf{63.9}} & 60.6 & 97.2 & \underline{\textbf{67.2}} & 53.3 & 97.8 & \underline{\textbf{62.0}}   \\ \cline{2-17}
        
        & 0.75 & 49.3 & 97.2 & 59.4 & 52.2 & 96.8 & \underline{\textbf{61.5}} & 54.9 & 96.9 & 63.7 & 60.5 & 97.1 & 66.9 & 52.9 & 98.0 & 61.6  \\ \cline{2-17}
        
        & 1 & 48.5 & 96.4 & 58.6 & 51.4 & 96.0 & 60.6 & 55.0 & 96.7 & 63.6 & 60.3 & 97.4 & 66.7 &  53.2 & 97.4 & 61.7  \\ \hline
        
        \multirow{3}{*}{\textbf{llama3}} & 0 & 46.5 & 95.9 & 56.6 & 51.4 & 95.5 & 60.5 & 55.0 & 96.2 & 63.6 & 57.4 & 96.9 & 63.7 & 52.9 & 98.0 & 61.6 \\ \cline{2-17}
        
        & 0.25 & 47.0 & 96.6 & \underline{\textbf{57.1}} & 51.6 & 96.2 & 60.8 & 55.0 & 96.2 & 63.6 & 56.8 & 96.9 & 63.4 & 53.0 & 97.7 & \underline{\textbf{61.7}} \\ \cline{2-17}
        
        \grayrow \cellcolor{white}
        & 0.5 & 45.9 & 96.2 & 56.2 & 51.6 & 95.8 & 60.6 & 55.3 & 96.4 & \underline{\textbf{63.8}} & 57.2 & 97.2 & \underline{\textbf{63.9}} & 52.9 & 98.0 & 61.6 \\ \cline{2-17}
        
        & 0.75 & 45.4 & 96.0 & 55.8 & 50.6 & 95.6 & 59.9 & 54.3 & 96.3 & 63.1 & 55.8 & 96.6 & 62.6 & 52.9 & 97.9 & 61.7 \\ \cline{2-17}
        
        & 1 & 46.3 & 96.9 & 56.6 & 52.1 & 96.4 & \underline{\textbf{61.2}} & 55.1 & 96.5 & 63.7 & 56.8 & 97.6 & 63.6 & 52.8 & 97.5 & 61.5 \\ \hline \hline
        \multicolumn{2}{l!{\vrule}}{\textbf{baseline (traditional analysis)}} & 26.1 & 97.9 & 34.8 & 45.1 & 97.4 & 53.4 & 49.2 & 97.4 & 57.5 & 31.4 & 99.4 & 38.2 & 52.9 & 98.0 & 61.6 \\ \hline
    \end{tabular}
}
    \label{tab:main_result}
\end{table*}

\subsection{Effectiveness of SEA~(RQ1)}~\label{sec:RQ1}


We evaluate the effectiveness of SEA when combined with FLTA, MLTA, and Kelp. 
The combinations work as follows: when combined with FLTA and MLTA, SEA is used to refine the callee set produced by these analyzers. 
For Kelp, cases with simple icalls, which already achieve 100\% accuracy, SEA directly use the results from Kelp. When Kelp falls back to FLTA or MLTA, SEA refines their results.
We conduct experiments across five temperature settings: 0, 0.25, 0.5, 0.75, and 1 with qwen and llama3.  
In our experiments, we find that SEA demonstrates optimal performance when using the qwen model at a temperature setting of 0.5. 
Consequently, we first present SEA's binary classification performance under this setting, where it predicts each call edge produced by the baseline as either a true or false edge.
The average results across 31 projects are shown in Table~\ref{tab:binary_result}.
Besides P, R, F, accuracy (A, correctly predicted call relations / total predicted call relations) is also presented.
We can observe that, compared to treating all call edges predicted by the baselines as true, SEA achieves a significant improvement in F1 score at the cost of approximately 1\% recall. Specifically, SEA enhances F1 by 14.5\% for Kelp, 17.1\% for MLTA, and 28\% for FLTA.
The diminishing performance improvement from FLTA to Kelp suggests that these baseline methods also prune some false targets, thereby reducing SEA's potential for enhancement. Furthermore, SEA's precision, at approximately 41\%, is partly influenced by the dataset. The presence of false negatives in the dataset significantly impacts the binary classification results.

To further understand the performance of SEA, we conduct a full evaluation and an exclusive evaluation.
For the full evaluation, the results represent the average icall analysis effectiveness across 31 projects. 
For the exclusive evaluation, we first categorize all icalls in the benchmark into three scenarios: FLTA-exclusive, MLTA-exclusive, and Kelp-exclusive. FLTA-exclusive denotes icalls analyzable only by FLTA, MLTA-exclusive denotes icalls refined by MLTA but not Kelp, and Kelp-exclusive denotes icalls refined by Kelp. Since Kelp already achieves 100\% accuracy, evaluating SEA's performance improvement on Kelp-exclusive cases is unnecessary. 
Therefore, we focus on its effectiveness in FLTA-exclusive and MLTA-exclusive scenarios. 
In the ground truth, these categories comprise 343, 373, and 82 cases, respectively. 
After filtering out invalid cases due to parsing errors where FLTA's analysis results do not include labeled true targets, we have 294 FLTA-exclusive and 359 MLTA-exclusive cases.
The exclusive evaluation results represent the average performance of SEA on all FLTA-exclusive and MLTA-exclusive cases.

The full and exclusive evaluation results are presented in Table~\ref{tab:main_result}. 
It is observed that the recall results of the baselines are less than 100\%, which is attributed to parsing errors in the code.
Through the analysis of the full evaluation results, we observe that when combined with FLTA, SEA demonstrates a significant performance gain. 
When using qwen as the main LLM, the F1 score increases from 34.8\% to between 58.5\% and 59.4\%, with a maximum recall loss of only 1.5\%. 
Although there is a slight decline in performance when using llama3 as the main LLM, we still observe an improvement in the F1 score of at least 21\%.
However, when combined with MLTA, the gain in SEA's F1 score decreases to 7.9\%-8.1\% using qwen and 6.5\%-7.8\% using llama3.
When combined with Kelp, the gain is only around 6\%.
\compactline

Through the analysis of the exclusive evaluation results, we can observe a significant improvement in SEA's performance in the FLTA-exclusive cases. 
When using qwen, the F1 score increases from 38.2\% to 66.4\%-67.2\%, while with llama3, the F1 score still rises to 62.6\%-63.7\%.
However, in the MLTA-exclusive cases, we observe that SEA brings almost no improvement, indicating SEA struggles to further optimize cases already refined by MLTA. 
We believe the underlying reason is that, in a certain sense, MLTA functions as a heuristic-based semantic analysis approach.

We illustrate this with Figure~\ref{fig:struct_example}.  
In the global, the function \texttt{ngx\_http\_log\_create\_main\_conf} is assigned to the \texttt{create\_main\_conf} field of \texttt{ngx\_http\_log\_module\_ctx}. 
In this example, MLTA successfully filters out a batch of false positives through the confinement between field \texttt{create\_main\_conf} of struct \texttt{ngx\_http\_log\_module\_ctx} and function \texttt{ngx\_http\_log\_create\_main\_conf}, while SEA eliminates the same false positives through the semantic similarity between them.
Therefore, SEA does not demonstrate a noticeable advantage in the MLTA-cases.

Regarding the FLTA-exclusive cases, We further categorizing the outcomes on these cases into three types: performance decrease, no change in performance, and performance improvement. 
Considering the potential inconsistency in LLM analysis results and qwen's superior performance compared to llama3, we focused on qwen's analysis. 
We find that during five times analysis, two cases experience a decrease in the F1 score, 61 cases show no change, and 206 cases consistently show an improvement in the F1 score.
The reason SEA can enhance FLTA's performance is that it leverages semantic information to filter out semantically irrelevant false positives. 
The motivating example in Figure~\ref{fig:icall_example} illustrates this well, where false targets are easily filtered out due to their complete functional dissimilarity with the call. 
SEA captures this semantic difference and filters out the false positive.
In addition, we also observe for a FLTA-exclusive case within the \texttt{cairo} project, the baseline approach identifies 208 callee targets, SEA successfully filter out the sole true target from this set.
Due to page limitations, more case studies can be found at our repository~\cite{CodeAnalyzer}.

For the 61 cases with no change in performance, we find that in 41 cases, FLTA alone achieves 100\% F1, indicating no room for further improvement. 
Regarding the remaining 21 cases, our manual analysis reveals that the callees filtered by FLTA had strong semantic similarities to the callers and were very likely to be invoked.
However, due to the complexity of the call relationships of the programs and considering the possibility of false negatives in the ground truth, we cannot be entirely certain about the presence or extent of false positives.
The failed cases on the one hand may be related to the hallucination issue~\cite{Hallucination} of LLMs. 
On the other hand, they can be attributed to ambiguous semantic information between the caller and the callee. 
For example, in a specific case in \texttt{vlc} project, the caller's context indicates that it is used to test the module's suitability, while the callee's context suggests it functions as an activate function, which is not related to the module. 
Consequently, the semantic analysis fails.
Nevertheless, we can still consider SEA to be quite effective in enhancing traditional methods.
\vspace{-0.3cm}
\noindent
\begin{tcolorbox}[size=title, opacityfill=0.1]
\textbf{ANSWER:} SEA substantially improves the effectiveness of traditional indirect call methods, increasing the F1 score by at least 20\% when combined with FLTA. 
SEA achieves a F1 score increase of up to nearly 30\% in the FLTA-exclusive cases. This improvement stems from its effective use of the LLM's semantic understanding to distinguish true/false callees.
\compactline
\end{tcolorbox}
\vspace{-0.3cm}

\subsection{Ablation Study~(RQ2)}

To further understand the contribution of local and global contexts to SEA, we analyzed this through three control groups:
\textbf{wo-local} -- using only the global context of the caller and callee;
\textbf{wo-global} -- using only the local context of the caller and callee;
and \textbf{wo-all} -- not using any local and global context, matching directly based on the caller statement and callee name.
\compactline

\begin{figure}[t]
  \centering
  \includegraphics[width=0.47\textwidth]{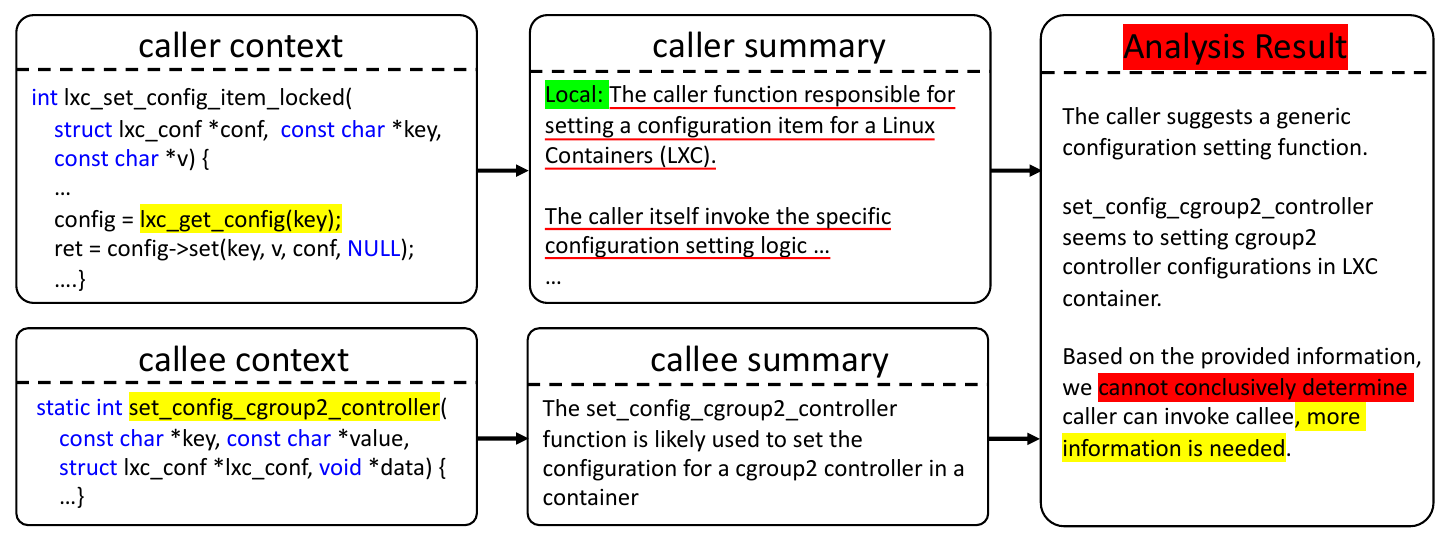}
  \vspace{-4mm}
  \caption{Example of removing local context.}
  \vspace{-4mm}
  \label{fig:ablation_example}
\end{figure}

The ablation study mostly follows the evaluation setup outlined in RQ1~(\S~\ref{sec:RQ1}) but only evaluates on the best temperature to save evaluation cost.
For qwen, the optimal value is 0.5. 
For llama3, the optimal temperature varies depending on the baseline combination. 
Notably, when combined with Kelp and analyzed exclusively in the FLTA cases, the best performance is observed at 0.5.
Therefore, we select this temperature setting for the ablation study.
Taking these findings into account, we conduct the ablation study using both qwen and llama3 at temperature of 0.5.
\compactline

\begin{figure}[t]
  \centering
  \includegraphics[width=0.47\textwidth]{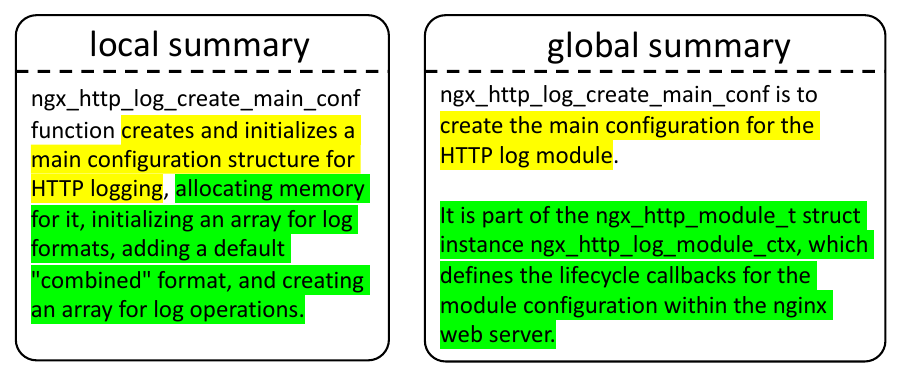}
  \vspace{-4mm}
  \caption{Similarities and differences of local/global contexts.}
  \vspace{-4mm}
  \label{fig:local_global}
\end{figure}

The experiment results are depicted in Table~\ref{tab:ablation_result}.
From this, we observed that for qwen, removing both local and global contexts significantly degrades SEA's performance. 
However, for llama3, its recall is much higher compared to qwen. 
Upon analyzing the results, we find that this difference is due to llama3's tendency to consistently answer \texttt{yes} when provided with insufficient information, while qwen exhibits more randomness in its responses.
In addition, the impact of eliminating global context is more significant than that of deleting local context, the performance degradation caused by removing the local context is less noticeable.
Upon contrasting this performance degradation with the results presented in Table~\ref{tab:main_result}, we observe a consistent and significant drop in performance across all tested temperature values.
This finding effectively mitigate the potential influence of randomness.

These findings indicate that both global and local contexts are essential for icall analysis. 
However, the comparison between the original and the wo-global group suggests that there is some overlap between global and local contexts. 
To further illustrate this issue, we analyze two specific cases. 
The first example, illustrated in Figure~\ref{fig:ablation_example}, highlights that in certain cases, the local context can effectively supplement the global context, ensuring a more accurate interpretation.
Here, the caller’s local context indicates its responsibility for setting configurations for a Linux container. 
When this local context is omitted during matching, the LLM interprets the function as a generic configuration setting function and subsequently indicates that more information is needed.
The second example, illustrated in Figure~\ref{fig:local_global}, presents the local and global summaries for the ``\url{ngx\_http\_log\_create\_main\_conf} '' function shown in Figure~\ref{fig:struct_example}. 
The yellow-highlighted parts indicate similarities between the local and global summaries, while the green-highlighted parts denote differences. 
Both summaries describe the function as creating the main configuration for the HTTP log module. 
However, the local summary provides a more detailed description of its specific functionality, whereas the global summary offers a higher-level description of the function's role within the lifecycle of an Nginx module. 
Because both the local and global summaries capture the core functionality of the function, the removal of either context in this case does not significantly impact SEA's performance.

\noindent
\begin{tcolorbox}[size=title, opacityfill=0.1, nobeforeafter]
\textbf{ANSWER:} Both local/global contexts contribute to SEA's performance, as they describe the caller and callee from different angles. However, there is a certain degree of overlap in the information they provide.
This indicates that when encountering parsing errors leading to partial information loss, SEA still has the potential to make correct analyses.
\compactline
\end{tcolorbox}

\begin{table}[t]
    \centering
    \caption{Ablation Study Under Temperature of 0.5.
    \vspace{-0.2cm}}
     \resizebox{0.9\linewidth}{!}{
    \begin{tabular}{c|c|c|c|c|c|c|c|c|c|c}
    \hline
    \multirow{2}{*}{\textbf{model}} & \multirow{2}{*}{\textbf{group}} &  \multicolumn{3}{c}{\textbf{FLTA}} & \multicolumn{3}{c}{\textbf{MLTA}} & \multicolumn{3}{c}{\textbf{Kelp}}
    \\ 
    \cline{3-11}
    & & P & R & F & P & R & F & P & R & F \\
    \hline
    \grayrow \cellcolor{white} 
    \multirow{3}{*}{\textbf{qwen}} & origin & 49.1 & 97.3 & \underline{\textbf{59.4}} & 51.9 & 96.9 & \underline{\textbf{61.4}} & 55.0 & 97.0 & \underline{\textbf{63.9}} \\ 
    \cline{2-11}
        
    & wo-local & 47.7 & 95.5 & 57.5 & 51.3 & 95.1 & 60.1 & 54.1 & 95.4 & 62.6 \\ \cline{2-11}
        
    & wo-global & 44.7 & 87.3 & 53.4 & 48.4 & 86.9 & 56.1 & 52.2 & 88.8 & 59.8 \\ \cline{2-11}
        
    & wo-all & 38.8 & 58.8 & 41.9 & 42.2 & 58.4 & 44.2 & 46.9 & 68.4 & 50.5 \\ \cline{2-11}

    \hline
        
    \grayrow  \cellcolor{white} 
    \multirow{3}{*}{\textbf{llama3}} & origin & 45.9 & 96.2 & \underline{\textbf{56.2}} & 51.6 & 95.8 & \underline{\textbf{60.6}} & 55.3 & 96.4 & \underline{\textbf{63.8}} \\ 
    \cline{2-11}
        
    & wo-local & 43.7 & 97.7 & 53.8 & 49.5 & 97.3 & 58.6 & 53.4 & 97.3 & 62.1 \\ 
    \cline{2-11}
        
        & wo-global & 43.7 & 89.1 & 52.6 & 49.4 & 88.7 & 57.1 & 52.5 & 90.2 & 60.3 \\ 
        \cline{2-11}
        
    & wo-all & 33.4 & 95.9 & 43.1 & 46.9 & 95.5 & 55.4 & 51.1 & 96.3 & 59.7  \\ 
        \hline
    \end{tabular}
     }
     \vspace{-10pt}
    \label{tab:ablation_result}

\end{table}

%% file: sections/5.discussion.tex
\section{Discussions}

\noindent
\textbf{Context Sensitivity} \tab
In this work, we follow previous studies to perform context-insensitive indirect call analysis. 
This approach can enhance tasks such as code navigation and improve bug detection and other inter-procedural analyses, similar to prior studies. 
However, unlike code navigation, a context-sensitive and more precise call analysis can further reduce false positives in static bug analysis. 
Therefore, an interesting future direction for inter-procedural bug detection would be to incorporate the function call chains involved in current indirect calls to perform a context-sensitive indirect call analysis.

\noindent
\textbf{Transferrability among Programming Languages}\tab Currently we only evaluated SEA in C language, in theory, our method can be adapted to other languages such as C++/Java/Python.
However, the diverse features of various languages may introduce new challenges to the application of SEA. 
For example, features like higher-order functions, reflection, and dynamic typing in Python can complicate the collection of global context, requiring adjustments to our context collection strategy. 
Therefore, adapting it to each language's unique features still require careful consideration.

\vspace{-5pt}
\section{Threats to Validity}

\noindent\textbf{Internal Validity}\tab
To measure the effectiveness of SEA, we followed previous works~\cite{cgpruner, autopruner, callee} by using dynamic execution to generate ground truth. 
However, due to inherent limitation of dynamic testing, the generated ground truth inevitably includes false negatives, and a more comprehensive large-scale manual labeling requires substantial effort. 
Conclusively, the biased dataset may negatively impact our experimental results.
To mitigate this concern, we supplemented our evaluation with manual analysis for certain cases.
\compactline

\noindent\textbf{External Validity}\tab External influences mainly arise from the following aspects: LLM selection, dataset, and compilation environment.
For LLM selection, we choose the open-source models Llama3 and Qwen1.5 for our experiments. However, as LLMs continue to evolve, both the open-source community and commercial platforms regularly update or release new models, which typically demonstrate enhanced capabilities. Consequently, some of the analysis results in this paper may need to be updated in the future.
Regarding the dataset, we select 31 projects from OSS-Fuzz as our benchmark. However, since these are relatively popular projects, they are likely to have been included in the training data of the LLMs. Therefore, it is unclear whether our experimental results can be generalized to the newest projects, especially those that are unlikely to appear in the training data.
For the compilation environment, we conduct our experiments exclusively in a no-compilation environment. The results demonstrate that SEA is highly effective for certain source code-level tasks. However, it remains unclear how much improvement our tool offers over traditional tools when extended to environments with compilation options.

%% file: sections/6.related_and_conclusion.tex
\vspace{-5pt}
\section{Related Work}\label{sec:related}

\subsection{Static Call Graph Analysis}

\noindent\textbf{Type Analysis For icall}\tab Type analysis~\cite{typeAnalysis, typro} considers all address-taken functions sharing identical function signatures with indirect calls as potential callees.
Due to its straightforward principles, it can be efficiently applied in control flow integrity schemes~\cite{EFCFI, TBCFI, FineCFI, modularCFI}, enabling scalability to millions of lines of code within minutes.
Subsequent works~\cite{Crix, FineCFI, MLTA}, while matching function signatures, concurrently consider the structure hierarchy of function pointers, leading to a substantial decrease in false positives.
Nevertheless, type analysis-based approaches continue to yield many false positives~\cite{ghavamnia2020temporal, Kelp}.
\compactline

\noindent\textbf{Pointer Analysis For icalls}\tab Pointer analysis~\cite{SVF, SUPA} treat function pointers as ordinary address-taken variables and aim to identify all address-taken functions aliased with them. 
Depending on the analysis dimensions, pointer analysis can be categorized into field-sensitive/insensitive, flow-sensitive/insensitive, and context-sensitive/insensitive, as well as inclusion-based and unification-based methods.
Generally, algorithms that consider more sensitive analyses tend to yield more accurate results but also incur higher performance overhead, leading to the ``pointer trap'' problem~\cite{pinpoint}. 
To mitigate performance overhead, on-demand pointer analysis approaches~\cite{pinpoint, sridharan2005demand, SUPA, sui2018value} have been proposed and have shown promising results. 
Nevertheless, scaling up existing precise demand-driven pointer analysis to hundreds of thousands of lines of code remains challenging.
In a recent study, conducted by Y. Cai et al. \cite{Kelp}, it was revealed that within the Linux kernel (version 5.15), 34.5\% of indirect calls are made via simple function pointers, while 23.9\% of address-taken functions are exclusively invoked through simple indirect calls. Consequently, they devised a regional def-use analysis approach named Kelp, which effectively diminishes erroneous callees by 54.2\%, while incurring only a negligible additional time cost of 8.5\%.
\compactline

\noindent\textbf{Deep Learning-Enhanced Call Graph Analysis}\tab In addition to these two lines of methodologies, Zhu et al.~\cite{callee} proposed a distinctive approach to icall analysis through similarity matching with a deep learning model, introducing a novel perspective and providing an alternative avenue for enhancing the precision of icall analyses.
In addition to icall analysis, another noteworthy area is call graph analysis in object-oriented languages such as Java. 
To reduce false positives produced by static analysis tools, Akshay et al.~\cite{cgpruner} proposed a method based on structural features. 
They extracted structural feature vectors from the call graph and then used a trained classifier to determine whether a caller-callee pair exists. 
Thanh et al.~\cite{autopruner} later extended this approach by incorporating the semantic features of the caller and callee functions' code into the structural features, thereby improving performance.
Although proven effective, these methods require labeled datasets for model training. However, acquiring a substantial dataset labeled with caller-callee pairs is exceedingly challenging for our task. Fortunately, the emergence of LLMs presents new opportunities to address this challenge.
\compactline

\subsection{LLM-assisted Program Analysis}

The emergence of LLMs has presented numerous new opportunities for program analysis. 
Over the past year, LLMs have made notable advancements in assisting fuzzing~\cite{TitanFuzz, FuzzGPT, KernelGPT}, static vulnerability detection~\cite{InferROI, gptscan}, and bug report analysis~\cite{LLift}. 
This is attributed to the powerful code understanding capabilities of LLMs, which reduce the dependency on heuristic rules in code analysis tools, thereby enhancing their performance.
Additionally, LLMs have been leveraged to augment various stages of the software development lifecycle. 
For Instance, LLMs have enhanced automated code generation~\cite{IDECoder} and refactoring tasks~\cite{pomian2024together}, streamlining software maintenance efforts. 
Overall, integrating LLMs into program analysis methodologies holds great potential for improving software systems' reliability, security, and efficiency.

\vspace{-5pt}
\section{Conclusion}

In this paper, we proposed SEA, a method that optimizes traditional static indirect call analysis through semantic analysis. 
By leveraging LLMs to understand the icall context, SEA effectively filters out semantically irrelevant caller-callee pairs from the static analysis results. 
Our experiments show that SEA refines the target set produced by static analysis methods through its code semantic understanding, which indicates a promising direction for semantic-enhanced program analysis.